# Local structures of gravity-free space and time and their phenomena

Jian-Miin Liu*
Department of Physics, Nanjing University
Nanjing, The People's Republic of China
*On leave. E-mail address: liu@phys.uri.edu

The progress in human civilization and human living style were undergoing acceleration in the last hundred years. Recalling the processes from Maxwell's theory of electromagnetic field to Hertz's verification experiment to nowadays communication and medium systems and those from the mathematical constructivity of computer pioneers Turing and von Newman to nowadays information high-way, data-base and data-treatment systems, we realize that the advances of science are the primary one of motive forces for this progress. Scientific researches, as being especially less commercial, deserve to receive more attentions and supports from human society. In which direction and how to support and invest in scientific researches ought to be perpetually an important issue for human society.

The Creativity Science Collaboration has singled out my work "Motivations to modify special relativity" [1] for the "Creative Scientist of the Year 2003" award winning work. In that work, I reported several observations on special relativity, its experimental facts and its relations to quantum mechanics and statistical mechanics. These observations made us conscious: Special relativity is not a ultimate theory; Some modification is needed; Any modification must not violate the principles of the constancy of the light speed and of the local Lorentz invariance; We probably have to change the assumption on local structures of gravity-free space and time in special relativity. Actually, such a kind of modification of special relativity has been already done. The generalized Finslerian structures of gravity-free space and time in the usual inertial coordinate system have been proposed. Here, in this lecture, I would like to talk about these structures of gravity-free space and time and their recognized phenomena.

Two fundamental postulates that Einstein states in his special theory of relativity are (i) the principle of relativity and (ii) the constancy of the speed of light in all inertial frames of reference. Conceptually, the principle of relativity implies that there exists a class of equivalent inertial frames of reference, any one of which moves with a non-zero constant velocity relative to any other and any one of which is supplied with motionless, rigid unit rods of equal length and motionless, synchronized clocks of equal running rate.

Einstein wrote: "in a given inertial frame of reference the coordinates mean the results of certain measurements with rigid (motionless) rods, a clock at rest relative to the inertial frame of reference defines a local time, and the local time at all points of space, indicated by synchronized clocks and taken together, give the time of this inertial frame of reference." [2] As defined by Einstein, in each inertial frame of reference, an observer can employ his motionless-rigid rods and motionless-synchronized clocks in the so-called "motionless-rigid rod and motionless-synchronized clock" measurement method to measure space and time intervals. By using this "motionless-rigid rod and motionless-synchronized clock" measurement method, the observer in each inertial frame of reference sets up his usual inertial coordinate system, denoted by $\{x^r, t\}$, $r=1,2,3$. Postulate (ii) asserts that the measured speed of light is the same constant c in every such usual inertial coordinate system.

Besides the two postulates, however, special relativity uses another assumption. This assumption concerns the Euclidean structure of gravity-free space and the homogeneity of gravity-free time in the usual inertial coordinate system,

$dX^2 = \delta_{rs} dx^r dx^s$, $r,s=1,2,3$, (1a)
$dT^2 = dt^2$, (1b)

anywhere and at any time.

Postulates (i) and (ii) and the assumption Eqs.(1a-1b) together yield the Lorentz transformation between any two usual inertial coordinate systems. Though this assumption was not explicitly articulated, most likely having been considered self-evident, Einstein said in 1907: "Since the propagation velocity of light in empty space is c with respect to both reference systems, the two equations, $x_1^2 + y_1^2 + z_1^2 - c^2 t_1^2 = 0$ and $x_2^2+y_2^2+z_2^2-c^2t_2^2=0$ must be equivalent." [3]. Leaving aside the question whether postulate (i) implies the linearity of transformation between any two usual inertial coordinate systems and the reciprocity of relative velocities between any two usual inertial coordinate systems, we know that the two equivalent equations, the linearity of transformation and the reciprocity of relative velocities exactly lead to Lorentz transformation.

As we pointed out in Ref.[1], an inconsistency exists in the theory formed by the two postulates and the assumption Eqs.(1a-1b). From the assumption Eqs.(1a-1b), one can find
$$Y^2 = \delta_{rs} y^r y^s, \tag{2}$$
where $Y=dX/dT$ is a real speed or velocity-length, and $y^r=dx^r/dt$, r=1,2,3, is the well-defined Newtonian velocity. Eq.(2) actually embodies the pre-relativistic velocity space,
$$dY^2 = \delta_{rs} dy^r dy^s. \tag{3}$$
This velocity space is characterized by boundlessness and the Galilean velocity addition law in the usual velocity-coordinates $\{y^r\}$, r=1,2,3. Conversely, special relativity owns the Lorentz transformation between any two usual inertial coordinate systems. That exactly indicates a finite velocity boundary at c and the Einstein law governing velocity additions.

Experimental facts clearly support the existence of the constancy of the light speed and the local Lorentz invariance. We have no choice but to change the assumption Eqs.(1a-1b). We assume that gravity-free space and time possess the following non-Euclidean structures in the usual inertial coordinate system $\{x^r, t\}$, r=1,2,3,
$$dX^2 = g_{rs}(dx^1, dx^2, dx^3, dt) dx^r dx^s, \quad r,s=1,2,3, \tag{4a}$$
$$dT^2 = g(dx^1, dx^2, dx^3, dt) dt^2, \tag{4b}$$
$$g_{rs}(dx^1, dx^2, dx^3, dt) = K^2(y)\delta_{rs}, \tag{4c}$$
$$g(dx^1, dx^2, dx^3, dt) = (1-y^2/c^2) \equiv g(y), \tag{4d}$$
$$K(y) = \frac{c}{2y}(1-y^2/c^2)^{1/2} \ell n \frac{c+y}{c-y}, \tag{4e}$$
$$y = (y^s y^s)^{1/2}, \quad y<c, \tag{4f}$$
$$y^s = dx^s/dt, \tag{4g}$$
where $dX$ and $dT$ are respectively the real space and time differentials between two neighboring points $(x^1, x^2, x^3, t)$ and $(x^1+dx^1, x^2+dx^2, x^3+dx^3, t+dt)$. The non-Euclidean structures of gravity-free space and time specified by two metric tensors $g_{rs}(dx^1, dx^2, dx^3, dt)$ and $g(dx^1, dx^2, dx^3, dt)$ are of the so-called generalized Finsler geometry. The generalized Finsler geometry is a generalization of Riemann geometry. It can be endowed with the Cartan connection. When and only when y approaches zero, metric tensors $g_{rs}(dx^1, dx^2, dx^3, dt)$ and $g(dx^1, dx^2, dx^3, dt)$ become the Euclidean.

The "motionless-rigid rod and motionless-synchronized clock" measurement method is not all that each inertial frame of reference has. For each inertial frame of reference, we imagine other measurement methods which are different from the "motionless-rigid rod and motionless-synchronized clock" measurement method. By taking these other measurement methods, an observer in each inertial frame of reference can set up other inertial coordinate systems, just as well as he can set up his usual inertial coordinate system by taking the "motionless-rigid rod and motionless-synchronized clock" measurement method. We call these other inertial coordinate systems the unusual inertial coordinate systems. One of the unusual inertial coordinate systems is the primed inertial coordinate system, denoted by $\{x'^r, t'\}$, r=1,2,3. The primed inertial coordinate system can be defined from the measurement point of view. We do not do this here.

We do believe in flatness of gravity-free space and time. We further assume that gravity-free space and time have the Euclidean structures in the primed inertial coordinate system,
$$dX^2 = \delta_{rs} dx'^r dx'^s, \quad r,s=1,2,3, \tag{5a}$$
$$dT^2 = dt'^2, \tag{5b}$$
anywhere and at any time.

The theory founded on two postulates (i) and (ii) and two assumptions Eqs.(4a-4g) and (5a-5b) has been formed, which is named the modified special relativity theory [4]. In the modified special relativity theory, the speed of light is the same constant c in all usual inertial coordinate systems and it is the localized Lorentz transformation, not Lorentz transformation, which stands as a linear transformation between any two usual inertial coordinate systems.

Dividing Eq.(4a) by Eq.(4b), we find

$$Y^2 = [\frac{c}{2y} \ln \frac{c+y}{c-y}]^2 \delta_{rs} y^r y^s, \quad r,s=1,2,3. \tag{6}$$

It has proved that Eq.(6) embodies the relativistic velocity space,

$$dY^2 = H_{rs}(y) dy^r dy^s, \quad r,s=1,2,3, \tag{7a}$$

$$H_{rs}(y) = c^2 \delta^{rs}/(c^2-y^2) + c^2 y^r y^s/(c^2-y^2)^2, \quad \text{real } y^r \text{ and } y<c, \tag{7b}$$

which has a finite boundary at c and the Einstein law for its velocity additions. The said inconsistency no longer exists.

We divide Eq.(5a) by Eq.(5b) for

$$Y^2 = \delta_{rs} y'^r y'^s, \quad r,s=1,2,3, \tag{8}$$

that embodies what the equation

$$dY^2 = \delta_{rs} dy'^r dy'^s, \quad r,s=1,2,3, \tag{9}$$

does, where $y'^s = dx'^s/dt'$, s=1,2,3, is called the primed velocity. Eqs.(8) and (9) represent the relativistic velocity space in the primed velocity-coordinates $\{y'^r\}$, r=1,2,3. In the primed velocity-coordinates, the relativistic velocity space is unbounded and its velocity additions obey the Galilean law.

We can find the relation between primed velocity $y'^r$ and Newtonian velocity $y^r$ from Eqs.(6) and (8), and also the relation between $dy'^r$ and $dy^r$ from Eqs.(7a-7b) and (9).

Maxwell, relying on two assumptions (a) the velocity distribution function is spherically symmetric and (b) the x-, y- and z- components of velocity are statistically independent, derived his non-relativistic equilibrium velocity distribution. Assumptions (a) and (b) reflect structural characteristics of the pre-relativistic velocity space in the usual velocity-coordinates, represented in Eq.(3). Now, for the relativistic velocity space, the Euclidean structure of it in the primed velocity-coordinates convinces us of Maxwell's distribution of primed velocities. From this Maxwell's distribution of primed velocities, using those relations that we find from Eqs.(6), (7a-7b), (8) and (9), we can obtain the relativistic equilibrium distribution of Newtonian velocities. It is

$$P(y^1,y^2,y^3) dy^1 dy^2 dy^3 = N \frac{(m/2\pi K_B T)^{3/2}}{(1-y^2/c^2)^2} \exp[-\frac{mc^2}{8K_B T}(\ln \frac{c+y}{c-y})^2] dy^1 dy^2 dy^3,$$

$$P(y) dy = \pi c^2 N \frac{(m/2\pi K_B T)^{3/2}}{(1-y^2/c^2)} (\ln \frac{c+y}{c-y})^2 \exp[-\frac{mc^2}{8K_B T}(\ln \frac{c+y}{c-y})^2] dy,$$

Where N is the number of particles, m is their rest mass, T is the temperature and $K_B$ is the Boltzmann constant. The relativistic equilibrium velocity distribution fits to the Maxwellian distribution for low-energy particles (y<<c) but substantially differs from the Maxwellian distribution for high-energy particles. It falls off to zero as y goes to c.

The relativistic equilibrium velocity distribution has been used to explain the observed non-Maxwellian decay mode of high-energy tails in velocity distributions of astrophysical plasma particles [5]. The deviation of the decay mode of these high-energy tails from Maxwellian has been observed for many years. Experimental data were mostly, if not all, modeled by the κ (kappa) distribution. As experimental data seem to be well modeled by the kappa distribution, as the kappa distribution contains a power-law decay when y goes to infinity, it was concluded that the decay mode of high-energy tails in velocity distributions of astrophysical plasma particles is power-law like. This conclusion is rather misleading. The kappa distribution shapes

$$K(y) dy = \frac{N}{\pi^{3/2}} \frac{1}{\theta^3} \frac{\Gamma(\kappa+1)}{\kappa^{3/2} \Gamma(\kappa-1/2)} (1+\frac{y^2}{\kappa \theta^2})^{-(\kappa+1)} dy$$

where $\theta = [(2\kappa-3)/\kappa]^{1/2}(K_B T/m)^{1/2}$, $\Gamma$ is the gamma function and kappa is a parameter to be determined in comparison with experimental data. Different values of kappa correspond to different kinds of velocity distribution. When and only when kappa goes to infinity, the kappa distribution becomes the Maxwellian. The kappa distribution can not be a good modeling distribution for experimental data. The reasons are: For any value of kappa, the kappa distribution extends as far as y=∞, while the velocity distribution involved in experimental data extends, we believe, as far as y=c; Velocity distribution of low-energy particles, as observed, can be well described by the Maxwellian distribution, but the kappa distribution with any finite kappa value does not reduce to the Maxwellian even for small y; the values of kappa in fitting experimental data vary from event to event. The relativistic equilibrium velocity distribution predicts a new decay mode for those high-energy tails: falling off to zero, as y goes to c, slower than any exponential decay, $\exp\{-[2c/(c-y)]^B\}$, and faster than any power-law decay, $(c-y)^n$, where B and n are two positive numbers.

The relativistic equilibrium velocity distribution has been also used to calculate the nuclear fusion reaction rate [5]. To create a nuclear fusion reaction, a proton or nucleus must penetrate the repulsive Coulomb barrier and be close to another proton or nucleus so that the strong interaction between them acts. The Coulomb barrier is in general much higher than thermal energy, their ratio is typically greater than a thousand. Nuclear fusion reactions can occur only among few high-energy protons and nuclei. If, under the conditions for nuclear fusion reactions, interacting protons and nuclei reach their equilibrium distribution in the period of time that is infinitesimal compared to the mean lifetime of nuclear fusion reactions, it is the equilibrium velocity distribution of these few high-energy protons and nuclei that participates in determining the rate of nuclear fusion reactions. In this instance, the difference between the Maxwellian velocity distribution and the relativistic equilibrium velocity distribution for high-energy protons and nuclei is non-negligible. Figure 1 sketches the situation.

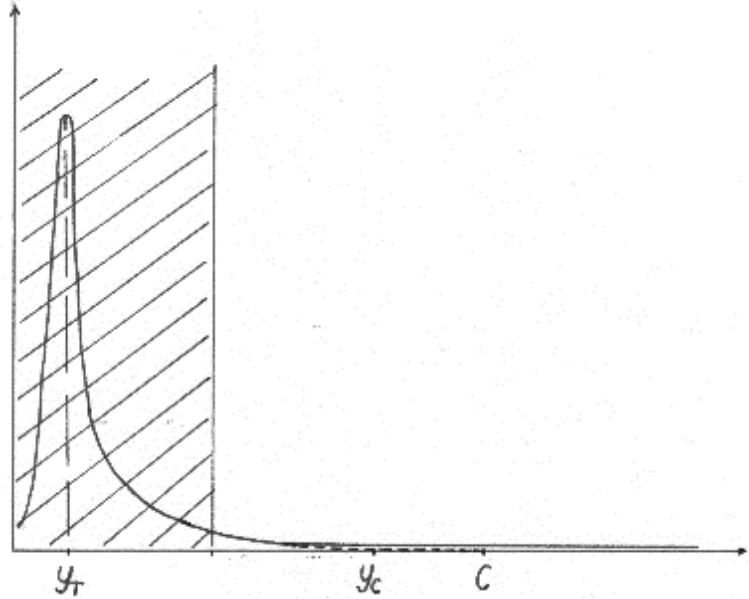

Fig.1 The Maxwellian velocity distribution (solid line) and the relativistic equilibrium velocity distribution (dash line) under the conditions in the Sun. As low-energy protons or nuclei in shadow area do not take part in solar nuclear fusion reactions, the difference between the Maxwellian velocity distribution and the relativistic equilibrium velocity distribution for high-energy protons or nuclei in un-shaded area becomes non-negligible.

In the situation, it is inappropriate to use the Maxwellian velocity distribution to calculate the nuclear fusion reaction rate. We have to use the relativistic equilibrium velocity distribution for the purpose. The calculation results indicate that the nuclear fusion reaction rate based on the relativistic equilibrium distribution, R, has a reduction factor with respect to that based on the Maxwellian velocity distribution, $R_M$,

$$R = \frac{\tanh Q}{Q} R_M,$$

$$Q = \left(2\pi z_1 z_2 \frac{K_B T}{\mu c^2} \frac{e^2}{\hbar c}\right)^{1/3},$$

where the reduction factor, tanhQ/Q, depends on temperature T, reduced mass μ and atomic numbers $z_1$ and $z_2$ of the studied nuclear fusion reactions. Since $0<Q<\infty$, the reduction factor takes values between 0 and 1, i.e. $0<\tanh Q/Q<1$. That gives rise to

$0<R<R_M$.

This inequality signifies much in resolving the solar neutrino problem. Solar neutrino problem is a long-standing puzzle in modern physics. Its main part is the discrepancies between the measured solar neutrino fluxes and those predicted by standard solar models. The measured solar neutrino fluxes range from 33%+/-5% to 58%+/-7% of the predicted values. We have examined standard solar models. All standard solar models take $R_M$ as a formula for the input parameters of solar nuclear fusion reaction rates. This is inappropriate. We should take R as a formula for the input parameters of solar nuclear fusion reaction rates. In doing so, standard solar models can be improved. The relativistic equilibrium velocity distribution is a possible solution to the solar neutrino problem [5].

More phenomena displaying the generalized Finslerian structures of gravity-free space and time, including those in gravity physics and quantum physics, are being to be recognized. In the presence of a gravitational field, according to Einstein's theory of gravitation, space and time co-join to be a four-dimensional space-time, and this space-time becomes curved. The line element of the curved space-time can be obtained by solving the Einstein field equation. The line element usually describes a Riemannian manifold. As a requirement, this line element must reduce to the Minkowskian when the gravitational field is removed or turned off. Now, based on the assumption Eqs.(4a-4g), some corrections to the obtained line element are necessary so as to make it reduce to the four-dimensional generalized Finslerian structure of gravity-free space-time when the gravitational field is removed or turned off. The corrected line element would represent a generalized Finslerian manifold, not the Riemannian manifold. The generalized Finsler geometry is a kind of generalization of Riemann geometry. There would be new physical phenomena involved in the generalized Finslerian manifold.

Thanks for your reading.